\documentstyle[times,graphics,epsfig,amsmath,amssymb,natbib]{mn2e}

\newcommand{\be}{\begin{equation}}
\newcommand{\e}{\end{equation}}
\newcommand{\bear}{\begin{eqnarray}}
\newcommand{\ear}{\end{eqnarray}}
\newcommand{\nline}{\nonumber \\}

\newcommand{\de}{{\rm d}}
\newcommand{\del}{\partial}

\def\apj{ApJ}

\def\mnras{MNRAS}

\def\u{{\vec U}}
\def\th{\vec{\theta}}
\def\rn{r_{\nu}}
\def\rnp{r'_{\nu}}

\def\E{\hat{E}}
\def\k{{\bf k}}
\def\kp{k_\parallel}
\def\HI{\rm HI}

\begin{document}
  \title[Detecting  ionized  bubbles ]{Detecting  ionized
 bubbles  in redshifted  21 cm maps}
\author[Datta,  Bharadwaj \& Choudhury  ]
{Kanan K. Datta$^1$\thanks{E-mail: kanan@cts.iitkgp.ernet.in}, 
Somnath Bharadwaj$^1$\thanks{E-mail: somnathb@iitkgp.ac.in}
and
T. Roy Choudhury$^2$\thanks{E-mail: chou@ast.cam.ac.uk}\\
$^1$Department of Physics and Meteorology \& 
Centre for Theoretical Studies, IIT, Kharagpur 721302, India\\
$^2$Institute of Astronomy, Madingley Road, Cambridge CB3 0HA, UK}

\maketitle
\date{\today}

\begin{abstract}
The reionization of the Universe, it is believed, occurred by the
growth of ionized regions (bubbles) in the neutral intergalactic
medium (IGM). We study the possibility of detecting these bubbles in
radio-interferometric observations of redshifted neutral hydrogen
(HI) $21$ cm radiation. The signal $(< 1 \, {\rm mJy})$ will be buried
in noise and foregrounds,  the latter being at least  a few orders of
magnitude stronger than the signal. We develop a visibility based
formalism that uses a filter to  optimally combine the entire signal
from a bubble while  minimizing the noise and foreground
contributions. This formalism makes definite predictions on the
ability to detect an  ionized bubble or conclusively rule out its 
presence in a radio-interferometric observation. We make predictions
for the currently functioning GMRT and a forthcoming instrument, the
MWA at a frequency of 150 MHz (corresponding to a redshift of 8.5).  
For both instruments, we show that a $3 \, \sigma$ detection 
will be possible  for a bubble  of comoving
radius $R_b \ge 40 \, {\rm Mpc}$ (assuming it to be spherical) in
$100 \, {\rm hrs}$ of observation and $R_b \ge 22 \, {\rm Mpc}$
in $1000 \, {\rm hrs}$ of observation, provided the bubble 
is  at the center of the field of view.  In both
these   cases the filter effectively removes the expected foreground 
contribution so  that it is below the signal, and the system noise is
the deciding  criteria. We find that there is a fundamental
limitation on the smallest bubble that can be detected arising from
the statistical fluctuations in the HI distribution. Assuming that
the HI traces the dark matter we
find that it will not be possible to detect bubbles with $R_b < 8 \,
{\rm Mpc}$ using the GMRT and  $R_b < 16 \,  {\rm Mpc}$ using the
MWA, however large  be the integration time.

\end{abstract}


\section{Introduction}
%Determining how and when the universe was reionized is an important
%issue. 
Quasar absorption spectra \citep{becker,fan} and CMBR
observations \citep{spergel,page} together imply that reionization
occurred over an extended period spanning the redshift range $6 \le z
\le 15$ (for reviews see \citealt{fck06,cf06}). It is
currently believed that ionized bubbles produced by the first luminous
objects grow and finally overlap to completely reionize the universe
\citep{barkana,furlanetto1}. In this paper we consider the possibility of detecting
these ionized bubbles in redshifted $21 \, {\rm cm}$ neutral hydrogen
(HI) maps.

An ionized bubble embedded in  neutral hydrogen will appear as a
decrement in the background  redshifted 21 cm radiation. This decrement
will typically span across several pixels and frequency channels in 
redshifted 21 cm maps. Detecting this is a big challenge
because the HI signal ($\sim 1 \,{\rm mJy}$ or lower ) will be
buried in foregrounds  \citep{shaver,dimat1,oh1,cooray3,santos05}
which are expected to be at least $2-3$ orders of magnitude larger. 
An objective detection criteria which optimally combines the  
entire signal in the  bubble while minimizing contributions from
foregrounds, system noise and other such sources 
is needed to search for ionized bubbles.   The noise in different
pixels of maps obtained from radio-interferometric observations is
correlated  (eg. \citet{thompson}), and  it is most convenient  to
deal with   visibilities instead.  These are the primary quantities
that are measured in radio-interferometry. In this paper we develop a
visibility based formalism to detect an ionized bubble
or conclusively rule it out in radio-interferometric observations of
HI at  high redshifts. 

The paper is motivated by the fact that the   Giant 
Metre-Wave Radio Telescope
 (GMRT\footnote{http://www.gmrt.ncra.tifr.res.in};  
\citealt{swarup}) which is currently functional 
has a band centered around  $150 \, {\rm MHz}$ which corresponds to HI
at $z\sim 8.5$. There are several low frequency radio telescopes which
are expected to become functional in the future  
(eg. MWA\footnote{http://www.haystack.mit.edu/arrays/MWA},
LOFAR\footnote{http://www.lofar.org/}, 21 CMA
\footnote{http://web.phys.cmu.edu/$\sim$past/} and 
SKA\footnote{http://www.skatelescope.org/}) all of which are being
designed to  be sensitive  to the epoch of reionization HI signal.
In this paper we apply our formalism for detecting ionized bubbles to make
predictions for the GMRT and for one of the forthcoming instruments,
namely the MWA. For both telescopes we investigate the feasibility of
detecting the bubbles, and in situations where a detection is
feasible we predict the required observation time. For both telescopes 
we make predictions for observations only at a single frequency ($150
{\rm MHz}$), the aim here being  to demonstrate the utility of our
formalism and not  present an exhaustive analysis of the feasibility of
detecting ionized bubbles in different scenarios and circumstances. 
For the GMRT we have used the telescope parameters from their
website, while for the MWA  we use the telescope parameters from 
\citet{bowman06}. It may be noted that MWA
is expected to be gradually expanded in  phases, and we have used
the parameters  corresponding to an early stage, the MWA - Low Frequency
Demonstrator. 

It is expected that detection of individual bubbles
would complement the studies of reionization 
through the global statistical signal of the redshifted $21 \, {\rm
  cm}$ radiation which has been studied extensively  (eg. \citealt{zal,
  morales04,bharad05a, bharad05}; for a recent review see
\citealt{furlanetto4}).   

The outline of the paper is as follows:
In Section 2 we discuss various sources which are expected to
contribute in low frequency radio-interferometric observation, this
includes the signal expected from an ionized bubble. In Section 3 we
present the formalism for detecting an ionized bubble, and in Section
4 we present the results and discuss its implications. 
The cosmological parameters used
throughout this paper  are  those determined as the best-fit values by
WMAP 3-year data release,  i.e.,  
$\Omega_m=0.23, \Omega_b h^2 = 0.022, n_s = 0.96, h = 0.74, \sigma_8
= 0.76$ (\citealt{spergel}).

\section{Different sources that contribute to low frequency radio
  observations} 

The quantity measured in radio-interferometric observations is the
visibility $V(\u,\nu)$ which is measured in a number of frequency
channels $\nu$ across a frequency bandwidth $B$ for every pair of
antennas 
in the array. For an antenna pair, it is convenient to use $\u={\vec
  d}/\lambda$ to quantify the antenna separation ${\vec d}$  projected
in the plane perpendicular to the line of sight    in units of the
observing wavelength $\lambda$. We refer to $\u$ as a baseline. 
The visibility 
is related to the specific intensity pattern on the sky $I_{\nu}(\th)$
as 
\be
V(\u,\nu)=\int d^2 \theta A(\th) I_{\nu}(\th)
e^{ 2\pi \imath \th \cdot \u}
\label{eq:1}
\e
where $\th$ is a two dimensional vector in the plane of the sky with
origin at the center of the field of view, and $A(\th)$ is the 
beam  pattern of the individual antenna.
 For the GMRT this can be well approximated by Gaussian
$A(\th)=e^{-{\theta}^2/{\theta_0}^2}$ where $\theta_0 \approx 0.6
~\theta_{\rm FWHM}$ and we use the values  $2.28\degr$  for $\theta_0$
  at $153 \, {\rm MHz}$ for the GMRT. Each MWA antenna element 
 consists of $16$  crossed dipoles distributed uniformly in a square
 shaped tile, and this is stationary with respect to the earth.
The MWA beam pattern is quite complicated, and it depends on the
pointing angle  relative to the zenith \citep{bowman07}. Our analysis
largely deals with the beam pattern within $1^{\circ}$ of the pointing
angle where it is reasonable to approximate the beam as being
circularly symmetric (Figures  3 and 5 of \citealt{bowman07} ). 
We  approximate the MWA antenna beam pattern as a Gaussian with
$\theta_0= 18\degr $  at $153 \, {\rm MHz}$.
Note that the MWA primary beam pattern is better
modeled as $A(\th)\propto \cos^2(K \theta)$, but  a Gaussian gives a
reasonable approximation in the center of the beam which is the region
of interest here. 
Equation (\ref{eq:1}) is valid
only under the assumption that the field of  view is small so that it
can be well approximated by a plane,  or under the unlikely
circumstances that all the  antennas are coplanar.

The visibility  recorded in $150 \, {\rm MHz}$ radio-interferometric
observations is a combination of three separate contributions 
\be
V(\vec{U},\nu)=S(\vec{U},\nu)+N(\vec{U},\nu)+F(\vec{U},\nu)
\label{eq:2}
\e
where $S(\vec{U},\nu)$ is the HI signal that we are interested in,
$N(\vec{U},\nu) $ is the system noise which is inherent to the
measurement and $F(\vec{U},\nu)$ is the contribution from other
astrophysical sources  referred to as the foregrounds. Man-made 
radio frequency interference (RFI) from cell phones and other
communication  devices are also expected to contribute to the measured
visibilities.
 Given the lack of a detailed model for the RFI
contribution, and anticipating that it may be possible to remove it
before the analysis, we do not take it into account here.    

\subsection{The HI signal from ionized bubbles}

According to models of reionization by UV sources, the early stages
of reionization are characterized by ionized HII regions around
individual source (QSOs or galaxies). As a first approximation, 
we consider these regions as ionized spherical bubbles characterized 
by three parameters, namely, its comoving radius $R_b$, the redshift
of its center $z_c$ and the position of the center determined by the
two-dimensional vector in the sky-plane $\th_c$.
The bubble is assumed to be embedded in an uniform intergalactic
medium (IGM)
with a neutral hydrogen fraction $x_{\rm HI}$. 
We use $\rn$ to denote the  comoving distance to the
redshift where the HI
emission, received at a frequency $\nu=1420 \, {\rm MHz}/(1+z)$,
originated, and define $\rnp=d \, \rn/d \, \nu$. The planar section
through the bubble at a comoving distance $\rn$ is a disk of comoving
radius $R_{\nu}=R_b \sqrt{1- (\Delta \nu/\Delta \nu_b)^2}$ where 
$\Delta \nu=\nu_c-\nu$ is the distance from the the bubble center $\nu_c$
in frequency space with $\nu_c=1420 \, {\rm MHz}/(1+z_c)$ and 
$\Delta \nu_b=R_b/r'_{\nu_c}$ is the bubble size in the 
frequency space.
The bubble, obviously, extends from 
$\nu_c-\Delta \nu_b$ to  $\nu_c+\Delta \nu_b$
in frequency and  in each frequency channel 
within this frequency range the image of the ionized bubble
is a circular disk of angular radius $\theta_{\nu}=R_{\nu}/r_{\nu}$;
the bubble  is  not seen in  HI  beyond this frequency range. 
Under such assumptions, the  specific intensity of the
redshifted HI emission is  
\be 
 I_{\nu}(\th)=\bar{I}_{\nu} x_{\HI}
\left[1-\Theta\left(1- \frac{\mid \th-\th_c\mid}{\theta_\nu} \right) \right] 
\Theta \left(1- \frac{\mid \nu -\nu_c \mid}{\Delta \nu_b} \right)
\label{eq:3}
\e
where $\bar{I_{\nu}}=2.5\times10^2\frac{Jy}{sr} \left (\frac{\Omega_b
  h^2}{0.02}\right )\left( \frac{0.7}{h} \right ) \left
(\frac{H_0}{H(z)} \right ) $  is the radiation background from the 
uniform HI distribution 
and $\Theta(x)$ is the Heaviside
step function. 

The soft X-ray emission from the quasar responsible for the ionized
region is expected to heat the neutral IGM in a shell around the
ionized bubble. The HI emission from this shell is expected to be
somewhat  higher than $\bar{I}_{\nu}$ \citep{wyithe04}. We do not
expect this to make a very big contribution, and we do not consider
this here. 

If we assume that the angular extent of
the ionized  bubble is  small compared to the angular scale of primary
beam {\it ie.} $\theta_\nu \ll \theta_0$,  we can take
$A(\th)$ outside the integral in 
eq. (\ref{eq:1}) and write the signal  as 
$A(\th_c) \int d^2 \theta I_{\nu}(\th)
e^{ 2\pi \imath \th \cdot \u}$, which essentially involves a
Fourier 
transform of the circular aperture $\Theta\left(1- \mid \th-\th_c\mid
r_{\nu}/R_{\nu}\right)$. For example, a bubble
of radius as large as 40 {\rm Mpc} at $z = 8.5$ would have an
angular size of only $\theta_\nu \approx 0.25^{\degr}$  which satisfies the
condition  $\theta_\nu  \ll  \theta_0$. 
In a situation where the bubble is at the center of the field of view, 
the visibility is found to be 
\be
S_{\rm center}(\u,\nu)=-\pi \bar{I_{\nu}} x_{\HI} \theta^2_\nu 
\left [ \frac{2 J_1(2 \pi U \theta_\nu
  )}{2 \pi U \theta_\nu}\right ] 
\Theta \left(1- \frac{\mid \nu -\nu_c \mid}{\Delta \nu_b} \right)
\label{eq:4}
\e
where $J_1(x)$ is the first order Bessel function. 
Note that $S_{\rm center}(\u,\nu)$  is real and it is the Fourier
transform of a circular 
aperture.  
The uniform HI background also contributes  
 $\bar{I_{\nu}} \pi \theta_0^2 {\rm e}^{-\pi^2 \theta_0^2 U^2}$ to the
visibility, but  this has been  dropped as it is quite insignificant
at the baselines 
of interest.  Note that the approximations used in eqs. (\ref{eq:4}) 
have been tested extensively by comparing the values with 
the numerical evaluation of the integral in eq. (\ref{eq:1}). 
We find that the two 
match to a high level of accuracy for the situations of interest
here. 
In the general situation where the bubble is shifted 
by $\th_c$ from the center of the field of
view, the visibility is given by
\be
S(\u,\nu) = {\rm e}^{-\theta_c^2/\theta_0^2} e^{2\pi i 
  \u \cdot \th_c}   S_{\rm center}(\u,\nu)
\label{eq:5}
\e
i.e., there is a phase shift of ${\rm e}^{2\pi i   \u \cdot \th_c}$ and 
a ${\rm e}^{-\theta_c^2/\theta_0^2}$  drop in 
the overall amplitude.

\begin{figure}
\includegraphics[width=0.45\textwidth]{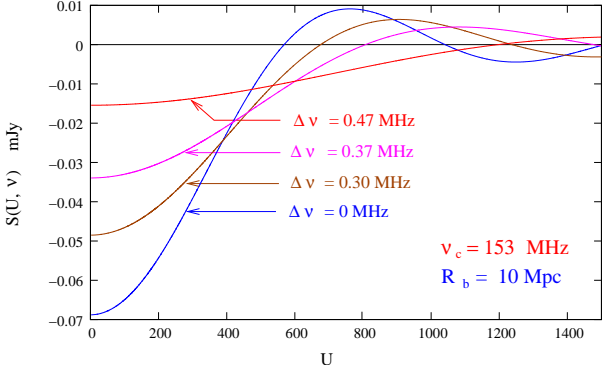}
\caption{Signal from a spherical ionized bubble of comoving
  radius $10 \, {\rm Mpc}$ as a function of baseline $U$ for
  different frequency channels.}  
\label{fig:bubu}
\end{figure}
\begin{figure}
\includegraphics[width=0.45\textwidth]{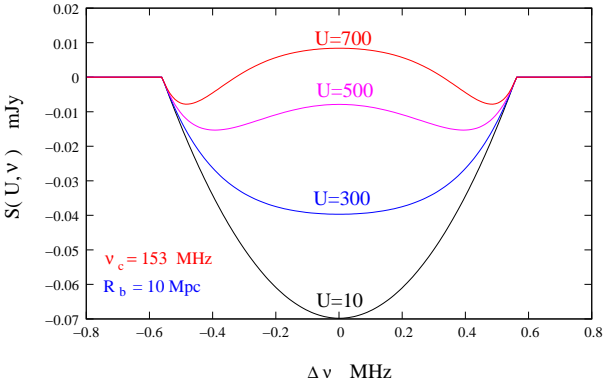}
\caption {Signal from a spherical  ionized bubble of comoving
  radius $10 \, {\rm Mpc}$ as a function of $\Delta \nu=\nu - \nu_c$   
 for different baselines.}
\label{fig:bubnu}
\end{figure}

Figures \ref{fig:bubu} and \ref{fig:bubnu} show the $U$ and $\Delta
\nu$ dependence of the visibility signal from an ionized
bubble with $R_b=10 \, {\rm Mpc}$ located at the center of the field
of view  at  $\nu_c= 153 \, {\rm MHz}$ ($z_c=8.3$),
assuming $x_{\rm HI}=1$. The signal extends over $\Delta \nu = \pm
\Delta \nu_b$  where $\Delta \nu_b=0.56 \, {\rm MHz}$. The extent in
frequency  $\Delta \nu_b = R_b/r'_{\nu_c}$ scales $\propto R_b$ when the
bubble size is varied. The Bessel function $J_1(x)$ has the first
zero crossing at $x=3.83$. As a result, 
the signal $S(\u,\nu)$ extends to $U_0=0.61
\rn [R_b \sqrt{1- (\Delta \nu/\Delta \nu_b)^2}]^{-1}$ where it has the
first zero crossing, and $U_0$  scales with the bubble size 
as $U_0 \propto 1/R_b$.  
The peak value of the signal is $S(0,\nu)=\pi
\bar{I}_{\nu} (R_b/\rn)^2  \sqrt{1- (\Delta \nu/\Delta \nu_b)^2}$ and
scales as $S(0,\nu) \propto R_b^2$ if the bubble size is
varied. We see that the peak value of the signal is $S(0,\nu_c)=70
\,{\rm   \mu Jy}$ for bubble size $R_b=10 \,{\rm Mpc}$ and would
increase  to  $1.75 \,{\rm  mJy}$ if $R_b=50 \,{\rm Mpc}$. 
Detecting these ionized bubbles will  be a big challenge because  the
signal is  buried in noise and foregrounds which are both
considerably larger in amplitude. 
Whether we are able to detect the ionized bubbles
or not  depends critically on our ability to construct optimal filters
which  discriminate  the signal from other contributions.    

\subsection{HI fluctuations}
In the previous sub-section, we assumed the ionized bubble
to be embedded in a perfectly uniform IGM. In reality, however, 
there would be fluctuations in the HI distribution in the IGM which,
in turn,
would contribute to the visibilities.
This contribution to the HI signal can be treated as a random variable
$\hat{S}(\u,\nu)$  with zero mean $\langle \hat{S}(\u,\nu) \rangle
=0$, whose  statistical properties are characterized by the
two-visibility correlation $\langle \hat{S}(\u_1,\nu_1)
\hat{S}(\u_2,\nu_2) \rangle$. This is related to $P_{\rm HI}({\bf k})$
the power spectrum of the 21 cm radiation efficiency in redshift space  
\citep{bharad04b} through 
\begin{eqnarray}
\langle \hat{S}(\u_1,\nu) \hat{S}^{*}(\u_2,\nu+\Delta \nu)\rangle 
\!\!\!\!\!&=&\!\!\!\!\!
\delta_{\u_1,\u_2} 
\frac{\bar{I}^2_{\nu} \theta_0^2}{2 \rn^2} \nonumber 
 \\ \!\!\!\!\!&\times  & \!\!\!\!\!\int_0^{\infty}
\de \kp \, P_{\rm HI}(\k)   
\cos(\kp \rnp \Delta \nu)
 \label{eq:a17} 
\end{eqnarray}
where $\delta_{\u_1,\u_2}$ is the Kronecker delta {\it ie.} different
baselines are uncorrelated,  To estimate the contribution from the HI
fluctuations we make the simplifying assumption 
that the HI  traces the dark matter,  which  gives 
$P_{\rm HI}(\k)=\bar{x}^2_{\HI}\left( 1+  \mu^2 \right)^2 
P(k)$ where $P(k)$ is the dark matter power spectrum and $\mu$ is the
cosine of the angle between $\k$ and the line of sight.  
This assumption is reasonable because the scales of interest are much
larger  than the Jeans length $\lambda_J\sim 10-100 \, \rm{kpc}$, and we
expect the HI to cluster in the same way as the dark matter.  

In addition to the above, there could be other contributions
to the HI signal too. For example, there would be 
several other ionized regions in the field of view
other than  the bubble under consideration.
The Poisson noise from these ionized patches  will increase the HI 
fluctuations and  there will also be an overall
drop in the contribution because of the reduced neutral fraction. 
These effects will  depend on the reionization model, and the
simple assumptions made in this paper would only provide a representative
estimate of the actual contribution.  Figure \ref{fig:fg} shows the
expected contribution from the HI fluctuations (HF) to the individual 
visibilities for GMRT and MWA.  
Note that while  this can be considerably larger than the signal that
we are trying to detect (particularly when the bubble size is  small), 
there is a big difference between the two.
The signal from the bubble is correlated across different
baselines and frequency channels whereas the contribution from random
HI fluctuations is uncorrelated at different baselines and it become
uncorrelated beyond a certain frequency  
separation $\Delta \nu$ \citep{bharad05a, datta}. 

\subsection{Noise and foregrounds}

\begin{figure*}
\includegraphics[width=0.9\textwidth]{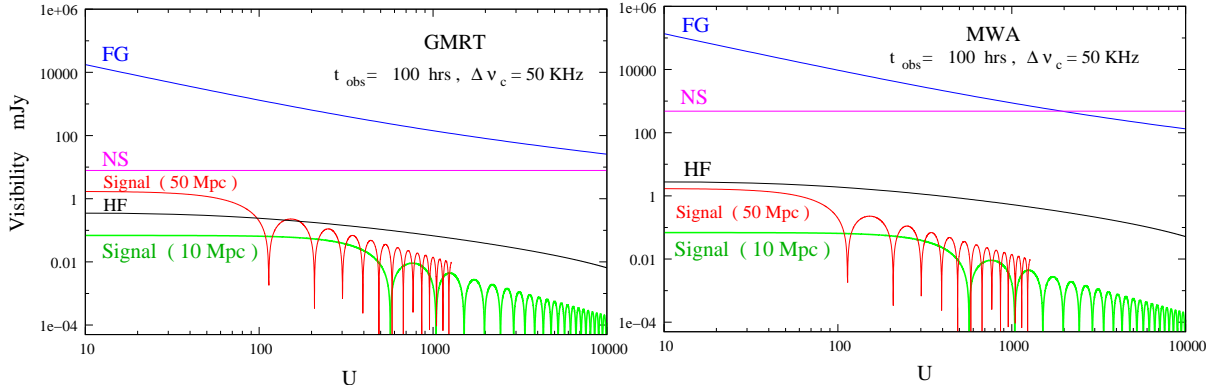}
\caption{The magnitude of the different contributions to the  visibility
 $V(\u,\nu)$   at $\nu=153 \,  {\rm  MHz}$ as a function of $U$. The 
signal,  foregrounds (FG), noise (NS) and HI fluctuations
(HF) contributions are shown for the GMRT (left) and MWA (right). The expected
  signal is shown  for  bubbles with radius $R=10 \, \rm{Mpc}$ and
 $R=50 \, \rm{Mpc}$.
The noise is estimated for a single baseline assuming an observation 
time $t_{obs}=100 \,
 {\rm  hrs}$ and channel width $\Delta \nu_c= 50 \, {\rm KHz}$.
 }
\label{fig:fg}
\end{figure*}

The system noise contribution $N(\u,\nu)$ in each baseline and
frequency channel is expected to be an independent Gaussian random
variable with zero mean ($\langle \hat{N} \rangle =0$) and  whose
variance is independent of   $\u$ 
and $\nu_c$. The predicted rms. noise contribution  is
(\citet{thompson})  
\be
\sqrt{\langle \hat{N}^2
  \rangle}=\frac{\sqrt2k_BT_{sys}}{A_{eff}\sqrt{\Delta \nu_c  
      \Delta t}}
\label{eq:rms}
\e
where $T_{sys}$ is the total system temperature, $k_B$ is the Boltzmann
constant, $A_{eff}$ is the effective collecting  area of each
antenna, $\Delta \nu_c$ is the channel width and $\Delta t$ is correlator
integration time. Equation (\ref{eq:rms}) can be rewritten as
\be
\sqrt{\langle \hat{N}^2  \rangle}
=C^x \left (\frac{\Delta \nu_c}{1
   \rm{MHz}} \right )^{-1/2}\left ( \frac{\Delta
    t}{1 \rm{sec}}\right )^{-1/2}
\label{eq:noise}
\e
where $C^x$ varies for different interferometric arrays. Using the
GMRT parameters 
$T_{sys}=482\rm{K}$ and $A_{eff}/2 k_B=0.33 \, {\rm{K/Jy}}$ at $153 {\rm
      MHz}$ gives $C^x=1.03 {\rm Jy}$ for the GMRT where as for MWA
  $T_{sys}=470\rm{K}$ and $A_{eff}/2 k_B=5\times10^{-3} \, {\rm{K/Jy}}$
  \citep{bowman06} gives $C^x=65.52{\rm Jy}$.
The rms noise  is reduced by a factor $\sqrt{\Delta t/t_{obs}}$ if we
average over $t_{obs}/\Delta t$ independent observations where
$t_{obs}$ is the total observation  time. 
Figure \ref{fig:fg} shows   the expected noise
for a single baseline at $153 {\rm MHz}$ for $\Delta \nu_c= 50 \, {\rm
  KHz}$ and  an observation  time of $100\,  {\rm hrs}$ for both the
 GMRT and MWA. Though $T_{sys}$ is nearly equal for the GMRT and the
 MWA, the noise in a single  baseline is   expected to be
$~60$ times larger for MWA than that for the GMRT. This is a
because the individual antennas have a much
 larger collecting area at the GMRT as compared to the MWA. 
 The fact that the MWA has
many more antennas $(N=500)$ as compared to the GMRT $(N=30)$ 
compensates for this. Note that nearly half (16) of the GMRT
antennas are at very large baselines which are not particularly
sensitive to the signal on the angular scales  the ionized bubble, and
only the other 14 antennas in the $1 \, {\rm km} \times 1 \, {\rm km}$
central square will contribute towards detecting the signal.  
For both the GMRT and the MWA, $T_{sys}$ is dominated by the sky
contribution $T_{sky}$ with the major contribution coming from our
Galaxy. We expect $T_{sys}$ to vary depending on whether the source is
in the Galactic plane or away from it. The value which we have used is
typical for directions off the Galactic plane. Further, the noise
contribution will also be baseline dependent which is not included
in our analysis.

Contributions from astrophysical foregrounds are expected 
to be several order of
magnitude stronger than the HI  signal. 
Extragalactic point sources and synchrotron radiation from our Galaxy 
are predicted to be the most dominant foreground components. Assuming
that the foregrounds    
are randomly distributed, with possible clustering, we have $\langle
\hat{F}(U,\nu) \rangle =0 $ for all the baselines other than 
the one at zero spacing ($U=0$), which is not considered
in this paper.
The  statistical properties are characterized by the
two-visibility correlation $\langle \hat{F}(\u_1,\nu_1)
\hat{F}(\u_2,\nu_2) \rangle$. We express this (details in
Appendix \ref{sec:vis})  in terms of  the multi-frequency angular
power spectrum 
(hereafter MAPS) $C_{l}(\nu_1\, 
\nu_2)$   of the brightness temperature fluctuations at the  
frequencies $\nu_1$ and $\nu_2$ as \citep{santos05,datta} 
\begin{eqnarray}
\langle \hat{F}(\u_1,\nu_1) \hat{F}(\u_2,\nu_2) \rangle
 =
\delta_{\u_1,-\u_2}  
\pi \left(\frac{   \theta_1^2 \theta_2^2}{\theta_1^2+
  \theta_2^2} \right )
\nline
\hspace{2cm} \left(\frac{\del B}{\del
  T}\right)_{\nu_1} \left(\frac{\del B}{\del T}\right)_{\nu_2} 
 C_{2  \pi   U_1}(\nu_1\, \nu_2).
\label{eq:11}
\end{eqnarray}
where $(\del B/\del  T)_{\nu}=2 k_B \nu^2/c^2$ is the
conversion factor to specific intensity, and we have assumed that the
primary beam pattern $A(\theta)=e^{-\theta^2/\theta_0^2}$ is frequency
dependent through $\theta_0 \propto \nu^{-1}$ and use $\theta_1$ and
$\theta_2$ to denote the value of $\theta_0$ at $\nu_1$ and $\nu_2$
respectively. Note that the foreground contribution to  different 
baselines are expected to be uncorrelated.

For each component of the foreground the MAPS is modeled as 
\be
C_{l}(\nu_1\, \nu_2)=A \left(\frac{\nu_f}{\nu_1} \right)^{ \bar{\alpha}}
\left(\frac{\nu_f}{\nu_2} \right)^{ \bar{\alpha}}
\left(\frac{1000}{l}\right)^{\beta}  I_l(\nu_1\, \nu_2).
\label{eq:fg}
\e
where  $\nu_f=130 \,{\rm  MHz}$,  and for each foreground
 component 
 $A$, $\beta$ and $\bar{\alpha}$
are the amplitude, the power law index of the angular power spectrum 
 and the mean spectral  index respectively.  The actual spectral  
index varies with line of sight across the sky and this causes the
foreground contribution to decorrelate with increasing frequency
separation $\Delta \nu=|\nu_1-\nu_2|$ which is quantified through the
foreground frequency  decorrelation function $I_l(\nu_1\, \nu_2)$
\citep{zal}  which has been  modeled as 
\be
I_l(\nu_1\, \nu_2)=\exp\left[ - \log_{10}^2 \left(\frac{\nu_2}{\nu_1}
 \right)/2   \xi^2 \right] \,.
\e
We consider the two most dominant  foreground
components namely extragalactic point sources and the 
diffuse synchrotron radiation from our own galaxy.  Point sources above
a flux level $S_{cut}$ can be identified in high-resolution continuum
images and removed. We note that absence of large baselines at the MWA
restricts the angular resolution, but it may be
possible to use the large frequency bandwidth $32 \, {\rm MHz}$ to
identify continuum point sources in the frequency domain. 
$S_{cut}$ depends on $\sigma$ the rms. noise  in the
image. We use   $S_{cut}=5 \sigma$ where $\sigma$ is the   
rms noise in the image given by (assuming 2 polarizations)
\be
\sigma=\frac{C^x}{\sqrt{2 N_b}}\left (\frac{B}{1
   \rm{MHz}} \right )^{-1/2}\left ( \frac{t_{obs}}{1 \rm{sec}}\right
)^{-1/2} 
\label{eq:12}
\e
where  $N_b=N(N-1)/2$ is the  number of  independent baselines, 
$N$ is the number of antennas in the array,
$B$ is the total frequency bandwidth and 
 $t_{obs}$ the  total observation time. For $t_{obs}=100 \, {\rm hrs}$
and $B= 6 \, {\rm MHz}$ we have $S_{cut}=0.1 {\rm mJy}$ for the GMRT and
using $B= 32 \,{\rm MHz}$ it gives $S_{cut}=0.2 \, {\rm  mJy}$  for
the MWA. 
 The value of $S_{cut}$ will be smaller for longer observations, but
reducing $S_{cut}$ any further does not make any difference to our
results so we hold $S_{cut}$ fixed at these values for the rest of our
analysis. The confusion noise from the unresolved
point sources is a combination of two parts, the Poisson contribution
due to the discrete nature of these sources and the clustering
contribution. The amplitude of  these two contributions have 
different $S_{cut}$ dependence. The parameter values that we have used
are listed in Table~\ref{tab:parm}. We have adopted the parameter
values from \citet{santos05} and incorporated the $S_{cut}$ dependence
from 
\citet{dimat1}.

\vspace{.2in}

\begin{table}
\caption{Parameters values used for characterizing different
  foreground contributions}
\label{tab:parm}
\begin{tabular}{|c|c|c|c|c|}
\hline
Foregrounds & $A ({\rm mK^2})$ & $\bar{\alpha}$ & $\beta$ & $\xi$\\ 
\hline
Galactic synchrotron & $700$ & $2.80$ &$2.4$ & $4$ \\
\hline
Point source & $61\left (\frac{S_{cut}}{0.1 {\rm mJy}}\right )^{0.5}$ 
& $2.07$ & $1.1$ &$2$ \\
(clustered part) & & & & \\
\hline
Point source & $0.16\left (\frac{S_{cut}}{0.1 {\rm
    mJy}}\right )^{1.25}$ & $2.07$ & $0$ & $1$ \\
(Poisson part) & & & & \\
\hline
\end{tabular}
\end{table}
\vspace{.2in}

Figure \ref{fig:fg} shows   the expected  foreground contributions 
 for the GMRT and MWA. 
The galactic synchrotron radiation is the most dominant foreground
 component  at large  angular scales ($U<1000$ for GMRT and $U<2000 $
 for MWA), while the clustering of the unresolved extragalactic point sources
 dominates at small angular scales. For all 
values of $U$, the foregrounds are at least four 
orders of magnitude larger than the signal, and also considerably
larger than the noise.

 The MWA has been designed with the detection of
 the statistical  HI fluctuation signal in mind, and hence  it
 is planned to have a very large field of view. 
 The foreground  contribution to a
 single baseline 
is expected to be $~10$ times stronger for the MWA than for the  GMRT 
because of a larger field of view.  As we shall show later, the
 increased foreground contribution is not a limitation for detecting
 HII bubbles. 
The foregrounds have a continuum spectra, and the contribution 
at two different frequencies at a  separation  $\Delta \nu$  are
expected to be 
highly correlated. For $\Delta \nu=1 \, {\rm MHz}$, the foreground
decorrelation function $I_l(\Delta \nu)$ falls by only $2 \times
10^{-6}$ for the galactic synchrotron radiation and by $3 \times
10^{-5}$ for the point sources. In contrast, the signal from an ionized
bubble peaks at a frequency corresponding to the bubble center and falls
rapidly with $\Delta \nu$ (Figure \ref{fig:bubnu}).
This holds the promise of allowing  the signal to be separated from
the foregrounds. 

\section{Formalism for detecting the ionized bubble}

We consider a radio-interferometric observation of 
duration $t_{obs}$,  carried out over the frequency range $\nu_1$ to
$\nu_2$.  
The HI signal from an ionized bubble, if it is present in the
data,  will be buried in foregrounds and
noise both of which are expected to be much larger. In this Section we
present a filtering technique aimed at detecting the signal from an
ionized bubble if it is present in our observations.
 To detect the signal from an ionized bubble of radius $R_b$ with
 center at redshift $z_c$ (or frequency $\nu_c$ )  and at an angle
 $\th_c$ from the center of the field of view, 
we introduce an estimator $\E[R_b,z_c,\th_c]$ defined as 
\be
\hat{E}=  \left[ \sum_{a,b} S_{f}^{\ast}(\u_a,\nu_b)
\hat{V}(\u_a,\nu_b) \right]/\left[   \sum_{a,b} 1 \right]
\label{eq:13}
\e
where $S_f(\u,\nu)$ is a filter which has been constructed to detect
the particular ionized bubble. 
Here $\u_a$ and $\nu_b$ refer to the
different baselines and frequency channels in our observations, and in
eq. (\ref{eq:13}) we are to sum over all independent data points
(visibilities).  
Note that the estimator $\E$ and the filter $S_f(\u,\nu)$ both depend
on $[R_b,z_c,\th_c]$, the parameters of the bubble we wish to
detect, but we do not show this explicitly. The values of these
parameters will be clear from the context.  

We shall be working in the continuum
limit where the two sums in eq.~(\ref{eq:13}) can be replaced by
integrals and we have 
\be
\E =\int d^2U \int d\nu  \, \rho_N(\u,\nu) \, \, 
{S_f}^{\ast}(\u,\nu) \hat{V}(\u,\nu)  
\label{eq:14}
\e
$d^2 Ud\nu \, \rho_N(\u,\nu)$ is the fraction  of data points {\it
  ie.} baselines and  frequency channels in the
interval  $d^2 U \, d \nu$. Note that $\rho_N(\u,\nu)$ is   usually
frequency   dependent,  and it is normalized so that $\int d^2 U \,\int
d\nu \, \rho_N(\u,\nu)=1$. We refer to $\rho_N(\u,\nu)$ as the
normalized baseline distribution function. 

We now calculate $\langle \E \rangle$ the expectation value of the
estimator. Here the angular brackets  denote an average with respect
different realizations of the HI fluctuations, noise and foregrounds,
all  of which have 
been   assumed to be random variables with zero mean. This gives
$\langle \hat{V}(\u,\nu) \rangle =S(\u,\nu)$ 
and 
\be
\langle \E \rangle  =\int d^2U \, \int d\nu  \, \rho_N(\u,\nu) \, \, 
{S_f}^{\ast}(\u,\nu) S(\u,\nu)  
\label{eq:15}
\e

We next calculate the variance of the estimator which is the sum of
the contributions  from the noise (NS), the foregrounds(FG)  and the
HI fluctuations (HF)
\begin{eqnarray}
\langle (\Delta \E)^2 \rangle &\equiv& \langle (\E- \langle \E \rangle
)^2\rangle 
\nonumber \\ 
&=&\left <(\Delta \hat
E)^2 \right >_{{\rm NS}}+\left<(\Delta \hat E)^2 \right >_{{\rm FG
}} \,
 +\left<(\Delta \hat E)^2 \right >_{{\rm HF}} \,.\nonumber \\ 
\label{eq:16}
\end{eqnarray}
To calculate the noise contribution we go back to eq. (\ref{eq:13})
and use the fact that the noise  in different baselines and
frequency channels are  uncorrelated.  We have 
\be
\langle (\Delta \hat E)^2 \rangle_{\rm NS}
= \langle \hat{N}^2  \rangle \left[
 \sum_{a,b} \mid S_{f}(\u_a,\nu_b)\mid^2 \right] / \left[ \sum_{a,b} 1
  \right]^2 
\label{eq:17}
\e
which in the continuum limit is 
\bear
\langle (\Delta \hat E)^2 \rangle_{\rm NS}
&=& \left[ \langle \hat{N}^2 \rangle /\sum_{a,b} 1 \right] 
\nline
&\times& \int d^2U \, \int d\nu  \, \rho_N(\u,\nu) \, \, 
\mid S_{f}(\u,\nu)\mid^2
\label{eq:18}
\ear
The term $ \sqrt{\left[ \langle \hat{N}^2 \rangle /\sum_{a,b} 1
  \right]}$  is the same as $\sigma$, the rms. noise in the image
  (eq. \ref{eq:12}). We then have  
\be
\langle (\Delta \hat E)^2 \rangle_{\rm NS}
= \sigma^2 
\int d^2U \, \int d\nu  \, \rho_N(\u,\nu) \, \, 
\mid S_{f}(\u,\nu)\mid^2\,.
\label{eq:20.a}
\e

For  the foreground contribution we have 
\bear
\left <(\Delta \hat E)^2 \right >_{\rm{FG}}\!\!\!\!\!&=&\!\!\!\!\!
\int\de^2 U_1 \int \de^2 U_2 \int \de \nu_1 \int \de \nu_2
\nline 
\!\!\!\!\!&\times&\!\!\!\!\!
\rho_N(\u_1,\nu_1) \rho_N(\u_2,\nu_2)
{S_f}^{\ast}(\u_1,\nu_1) {S_f}^{\ast}(\u_2,\nu_2)
\nline 
\!\!\!\!\!&\times&\!\!\!\!\!
\langle \hat{F}(\u_1,\nu_1) \hat{F}(\u_2,\nu_2) \rangle 
\ear
In the continuum limit we have  (details given in 
Appendix \ref{sec:vis})
\bear
\langle \hat{F}(\u_1,\nu_1) \hat{F}(\u_2,\nu_2) \rangle &=&
\delta_D^{(2)}(\u_1+\u_2)  \left(\frac{\del B}{\del  T}\right)_{\nu_1}
\left(\frac{\del B}{\del  T}\right)_{\nu_2} 
\nline
&\times&
C_{2 \pi U_1}(\nu_1,\nu_2)
\label{eq:21}
\ear
which gives the variance of the foreground contribution to be 
\bear
\left <(\Delta \hat E)^2 \right >_{\rm{FG}}\!\!\!\!\!&=&\!\!\!\!\!
\int\de^2 U  \int \de \nu_1 \int \de \nu_2 \left(\frac{\del B}{\del  T}\right)_{\nu_1} \left(\frac{\del B}{\del
  T}\right)_{\nu_2} 
 \nline 
\!\!\!\!\!&\times&\!\!\!\!\!
\rho_N(\u,\nu_1) \rho_N(\u,\nu_2)
{S_f}^{\ast}(\u,\nu_1) {S_f}(\u,\nu_2)
 \nline 
\!\!\!\!\!&\times&\!\!\!\!\!
 C_{2 \pi U}(\nu_1,\nu_2)
\label{eq:fg1} 
\ear

We use eq. (\ref{eq:fg1}) to calculate  
$\left <(\Delta\hat E)^2 \right >_{\rm{HF}}$  too, 
 with the difference that we use 
the power spectrum 
$C_{2 \pi   U}(\nu,\nu + \Delta \nu)$  for the HI fluctuation  from
\citet{datta} instead of the foreground contribution. 

In an observation it will be possible to detect 
the presence of an ionized bubble having parameters $[R_b,z_c,\th_c]$ 
at, say 3-sigma confidence level,
if $\langle \hat{E} \rangle \ge 3 \sqrt{\langle (\Delta \hat{E})^2
  \rangle}$. In such a situation, an observed value $E_o$ can be
interpreted as a detection with  $99.7 \%$ (i.e., 3-sigma)
confidence  if $E_o > 3 \sqrt{\langle (\Delta \hat{E})^2
  \rangle}$. The presence of the ionized bubble can be ruled out at
the same level  of confidence if $\langle \hat{E} \rangle - E_o > ~3~
\sqrt{\langle (\Delta  \hat{E})^2   \rangle} $. 

\subsection{Baseline distribution}
\begin{figure}
\includegraphics[width=0.45\textwidth]{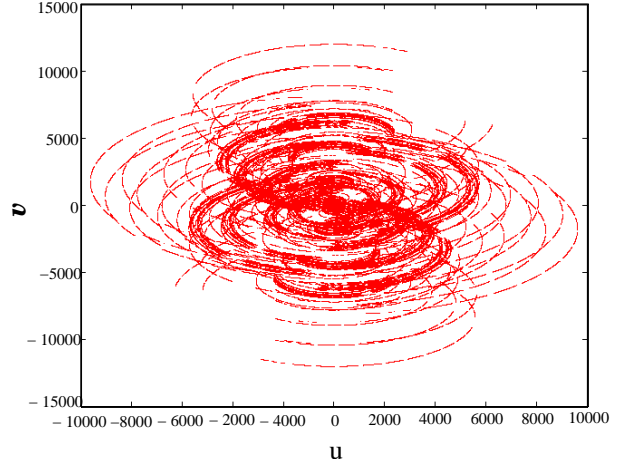}
\caption{This shows the baseline coverage for  $14 \, {\rm hrs}$ of 
 GMRT  $153 \, {\rm MHz}$ observation  at $ 45 \degr$ declination.}
\label{fig:4}
\end{figure}
\begin{figure}
\includegraphics[width=0.45\textwidth]{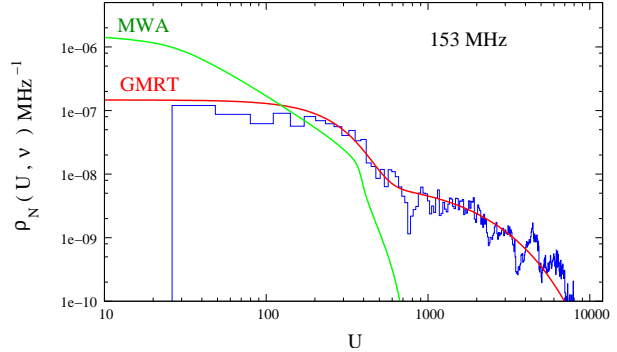}
\caption{This shows the normalized baseline distribution
  $\rho_N(U,\nu)$ for the GMRT and the MWA at $153 \, {\rm MHz}$. The
  wiggly curve  shows the actual values for the GMRT observation shown
  in Figure \ref{fig:4} and  the smooth curve is the analytic fit.} 
\label{fig:5}
\end{figure}

In this subsection we discuss the normalized baseline distribution
function $\rho_N(\u,\nu)$ which has been introduced earlier. 
Figure \ref{fig:4} shows the baseline 
coverage for $14 \, {\rm hrs}$ of observation towards a
 region at  declination $\delta=45\degr$  with the GMRT  at
$153 {\rm MHz}$. In this figure  $u$ and $v$ refer to the Cartesian
 components of the baselines $\u$. Note that the baseline distribution
 is not exactly circularly symmetric. This asymmetry depends on the
source  declination which would be different for every observation. 
 We make the simplifying assumption that the baseline  distribution is
 circularly symmetric whereby  $\rho_N(\u,\nu)$ is a function of $U$. 
This considerably simplifies our analysis and gives reasonable 
estimates of what we would expect over a range of declinations.  
Figure \ref{fig:5} shows $\rho_N(\u,\nu)$ for the GM RT determined from
the baseline coverage shown in Figure  \ref{fig:4}. We find that this
is well described by the sum of a Gaussian and an exponential
distribution. The GMRT has a hybrid antenna distribution \citep{chengalur}
with $14$ antennas being randomly distributed in a central square
approximately $1 \, {\rm km} \times 1 \, {\rm km}$ and $16$ antennas being
distributed along a Y each of whose arms is $14 \, {\rm km}$ long. 
The Gaussian gives a good fit at small baselines in 
the central  square and the exponential fits the large baselines.  
Determining the best fit parameters using a least square gives 
\bear
\rho_N(\u,\nu)&=&\frac{1}{B}\left(\frac{\lambda}{{1 \, \rm km}}\right)^2 \left [ 0.21
 \exp \left ( 
  -\frac{{U \lambda}}{2 a^2}\right ) \right.
\nline
&+& \left. 9.70 \times 10^{-3} \exp
  \left ( -\frac{U \lambda-b}{d}\right )\right ]
\label{eq:22}
\ear
where  $a=0.382 \, {\rm km}$, $b=0.986 \, {\rm km}$, $d=3.07 \, {\rm
  km}$ and $B$ is the frequency bandwidth which has a maximum value of
$6 \, {\rm MHz}$. 

Following \citet{bowman06} we assume that the MWA  antennas are
distributed within a radius of $0.750 \, {\rm km}$  with  the density of
antennas decreasing with radius $r$ as $\rho_{ant}(r)\propto
  r^{-2}$ and  with a  maximum density of one antenna per $18 {\rm
    m^2}$. The
normalized baseline distribution is  estimated in terms of 
$\rho_{ant}(r)$ and we have 
\bear
\rho_N(\u,\nu)&=&\frac{1}{4.4\times10^2 }\frac{1}{B}\left(\frac{ \lambda }{1 \,{\rm
    km}}\right)^2 \int_{
    r=0}^{\infty}\de^2r\rho_{ant}(
  r)
\nline
&\times&
\int_{\phi=0}^{2\pi}\rho_{ant}(|{\vec{r}} -\lambda \, \u \,|)\de \phi
\label{eq:23}
\ear
where the bandwidth $B$ is $32\, {\rm MHz}$, $|{\vec{r} }-\lambda \,
\u | = \left (  r^2 + U ^2\,\lambda ^2 - 
2  r\, \lambda \,
  U \, cos\phi \right )^{1/2}$. Note that $\rho_N(\u,\nu) $ depends on
the observed frequency.
Figure \ref{fig:5} shows the normalized baseline distribution function
$\rho_N(\u,\nu)$ for both  the GMRT and the MWA. We see that maximum
baseline for the GMRT is $U_{max}\sim  10,000$ whereas $U_{max}\sim
750$ for the MWA. However, the smaller baselines will be sampled more densely 
in the MWA as compared to the GMRT. 

\subsection{Filter}

\begin{figure*}
\includegraphics[width=0.90\textwidth]{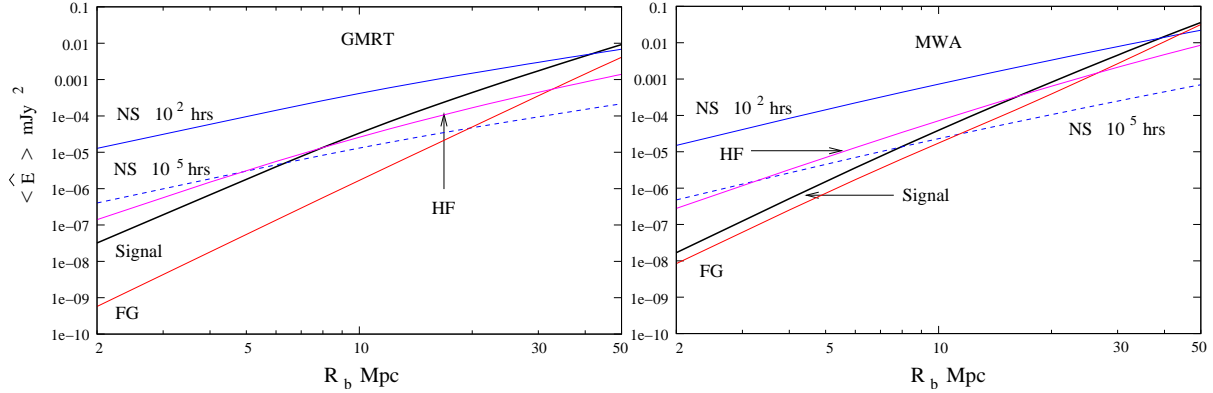}
\caption{The signal  quantified through the  expectation value of the 
  estimator $\langle \hat{E} \rangle$ for Filter I. The 
other components  (NS - Noise, FG - Foregrounds, HF - HI Fluctuations)
  are quantified through their   contribution to the 3-sigma fluctuation  
  $3 \times \sqrt{\langle  ( \Delta \hat{E}  )^2 \rangle}$.  
}
\label{fig:int_fil0}
\end{figure*}

It is a major challenge  to detect the signal which is expected to be
buried in noise and foregrounds both of which are much stronger
(Figure \ref{fig:fg}). It would be relatively simple to detect the
signal in a situation where there is only noise and no
foregrounds. The signal to noise ratio (SNR) is maximum  if we use
the  signal that we wish to detect as the filter ($ie. \, 
S_f(\u,\nu)=S(\u,\nu)$) and the SNR  has a value  
\bear
\frac{\langle ( \hat E) \rangle }{\sqrt{\langle (\Delta \hat
E)^2 \rangle_{\rm NS}}}\!\!\!\!\!&=&\!\!\!\!\!
\frac{1}{\sigma}
\left[ \int \, \de^2 U \, \int \de \nu \, \rho_{N}(\u,\nu) \mid
  S(\u,\nu) \mid ^2  \right]^{0.5}
\nline
\!\!\!\!\!&\propto&\!\!\!\!\! \sqrt{t_{obs}}\,.
\label{eq:s1}
\ear
The observing time necessary for a $3$-$\sigma$
detection (i.e., SNR $=3$) would be the least for
this filter. The difficulty with 
using this filter is that  the foreground contribution to
$\sqrt{\langle   (\Delta \hat  E)^2 \rangle}$ is orders of magnitude
more than   $\langle ( \hat E) \rangle$.  The foregrounds, unlike the
HI signal,  are all expected to have a smooth frequency dependence and
one requires filters which incorporate this fact so as to reduce
the foreground contribution. We
consider two different filters which  reduce
the foreground contribution, but it occurs at the expense of
reducing the SNR, and  $t_{obs}$ would be more than that
predicted by eq. (\ref{eq:s1}).

The first filter (Filter I) subtracts out any frequency independent
component from 
the frequency range $\nu_c-B^{'}/2$ to $\nu_c+B^{'}/2$ with $B{'} \le B$
{\it ie.} 
\bear
S_f(\u, \,\nu)\!\!\!\!\!&=&\!\!\!\!\!\left(\frac{\lambda_c}{\lambda}\right)^2\left[ S(\u, \,\nu)\right.
\nline
\!\!\!\!\!&-&\!\!\!\!\! \left. \frac{\Theta(1-2 \mid \nu -\nu_c \mid/B^{'}) }{ B^{'}}
\int_{\nu_c-B^{'}/2}^{\nu_c + B^{'}/2}S(\u,\nu') \, \de \nu' \right ].
\nonumber\\ 
\label{eq:25}
\ear
This filter has the advantage that it does not require  any prior
knowledge about the foregrounds except that they have a continuous
spectrum. It has the drawback that there will be contributions from the
residual foregrounds as all the foregrounds are expected to have a
power law spectral dependence and not a constant. A larger value of
$B^{'}$  causes the SNR to 
increases,  and in the limit $B^{'} \rightarrow {\infty}$
the SNR  approaches the value given in
eq. (\ref{eq:s1}). Unfortunately the residues in the foregrounds also
increase with $B^{'}$. We use $B^{'}=4 \Delta \nu_b$ provided it is
less than $B $, and $B^{'}=B $ otherwise.  

The  frequency dependence of the total foreground contribution
can be expanded in Taylor series. Retaining terms only up to the first order
we have 
\be
C_l(\nu_1,\,\nu_2)=  C_l(\nu_c,\,\nu_c)
\left [1- \left (\Delta \nu_1+\Delta
  \nu_2 \right )\alpha_{eff} /\nu_c \right]
\label{eq:24}
\e
where  $\Delta \nu=\nu -\nu_c$ and $\alpha_{eff}=\frac{\sum_i \alpha^i
  \, A^i \left (1000/l\right)^{\beta_i}}{\sum_iA^i \left ( 
1000/l\right )^{\beta_i}}$ is the   effective spectral index,
here $i$ refers to the different foreground components. Note that
$\alpha_{eff}$ is $l$ dependent.  The second filter that we consider
(Filter II) allows for a linear frequency dependence of the
foregrounds and we have 
\bear
S_f(\u,\nu)\!\!\!\!\!&=&\!\!\!\!\!(1+\alpha_{eff}\Delta \nu /\nu_c)\left(\frac{\lambda_c}{\lambda}\right)^2 \left[
S(\u,
\,\nu) \right.
\nline
\!\!\!\!\!&-&\!\!\!\!\!\left.
\frac{\Theta(1-2 \mid \nu -\nu_c \mid/B^{'}) }{ B^{'}}
\int_{\nu_c-B^{'}/2}^{\nu_c + B^{'}/2}
S(\u,\nu') \, \de \nu'  \right].
\nonumber\\ 
\label{eq:26}
\ear
Note that for both the filters we   include an extra factor 
$\left(\lambda_c/\lambda\right)^2$. 
%which has not been shown in
%equations (\ref{eq:25}) and (\ref{eq:26}).  
This is introduced with
the purpose of canceling out the $\lambda^2$ dependence of
of the normalized baseline distribution function  $\rho_N(\u,\nu)$ and 
this  substantially reduces the foreground contribution.  

\section{Results and Discussions}

\begin{figure*}
\includegraphics[width=0.90\textwidth]{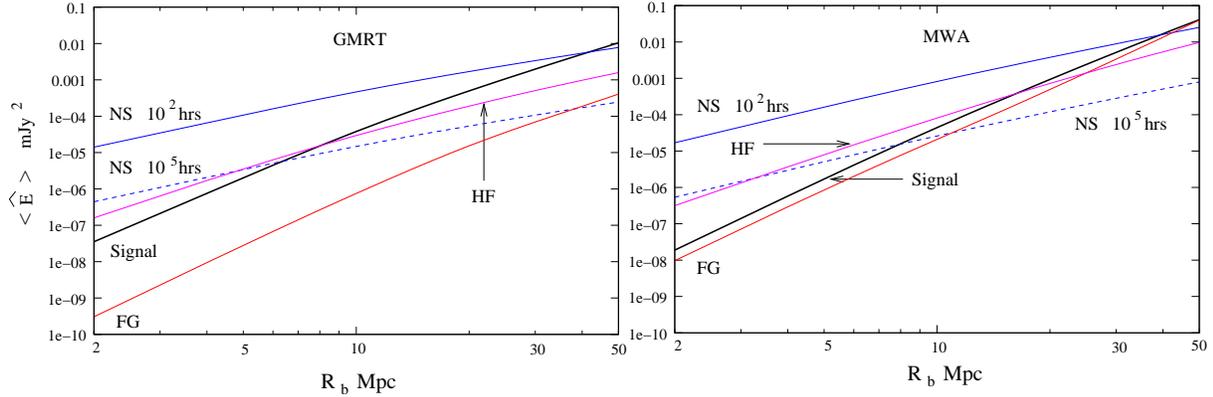}
\caption{Same as Figure \ref{fig:int_fil0} except that 
  Filter II is used instead of Filter I.}
\label{fig:alpha_eff}
\end{figure*}

We first consider the most optimistic  situation where the bubble is
at the center of the field of view and the filter center is exactly
matched with the bubble center. 
The size distribution of HII regions are
quite uncertain, and  would depend on the reionization history and the
distribution of ionizing sources. However, there are some indications in
the literature on what could be the typical size of HII regions. For
example, \citet{wyithe05} deduce from proximity zone
effects that $R_b \approx 35$\,\rm{Mpc} at $z\approx 6$, which should be
considered as a lower limit. On the other hand,
\citet{furlanetto05} (Figure 1(a))
infer that the characteristic bubble size
$R_b>10 \,\rm{Mpc}$ at $z=8$ if the ionized fraction $x_i>0.75$ ($R_b \sim 50
\rm{Mpc}$ if $x_i \sim 0.9$). Theoretical models which match a variety of 
observations \citep{choudhury07} imply that $x_i$ could be as high
as $90 \, \%$ at $z \sim 8$, which would mean bubble sizes of
$\,\sim40-50$\,\rm{Mpc}. To allow for the large  variety of possibilities,   
 we have presented  results for a wide range of $R_b$ values from $2
\, {\rm Mpc}$ to $50 \, {\rm Mpc}$. We
restrict our analysis to a situation where the IGM  outside the bubble 
is completely neutral ($x_{\rm HI}=1$).  The signal would fall
proportional to 
$ x_{\rm HI}$ if the IGM outside the bubble were partially
ionized  ($x_{\rm HI}<1$).  
The expected signal  
$\langle \hat{E} \rangle$ and 3-sigma fluctuation $3 \times
\sqrt{\langle  ( \Delta \hat{E}  )^2 \rangle}$ from each of the
  different components discussed in Sections 2 and 3 as a function 
of bubble size $R_b$ are shown in Figures \ref{fig:int_fil0} 
and \ref{fig:alpha_eff}.
Both the figures show exactly the
same quantities, the only difference being that  they refer to Filter
I and Filter II respectively. 
A  detection is possible only in situations where $\langle \hat{E} 
  \rangle > 3 \times \sqrt{\langle  ( \Delta \hat{E}  )^2 \rangle}$, 
 the {\it rhs. } now refers to the total contribution to the
estimator variance 
from all the components. 

The signal is expected to scale as $R_b^3$ and the noise as  $R_b^{3/2}$ 
in a situation where the baseline distribution is uniform {\it ie.}
$\rho_N(\u,\nu)$ is independent of $U$.  This holds at $U <300$ for
the GMRT (Figure \ref{fig:5}), and the expected scaling is seen for
$R_b \ge 20 \, {\rm 
  Mpc}$. For smaller bubbles the signal extends to larger baselines
where $\rho_N(U,\nu)$ falls sharply, and the signal and the noise
both have a steeper $R_b$ dependence. The MWA baseline distribution is
flat for only a small $U$ range   (Figure \ref{fig:5}) beyond which it
drops. In this case the signal and noise are found to scale as $R_b^4$
and $R_b^2$ 
respectively. Note that the maximum baseline at  MWA is $U=750$, and 
hence a considerable amount of the signal is lost for  $R_b < 10 \, {\rm
  Mpc}$. 

 \begin{figure}
\includegraphics[width=0.45\textwidth]{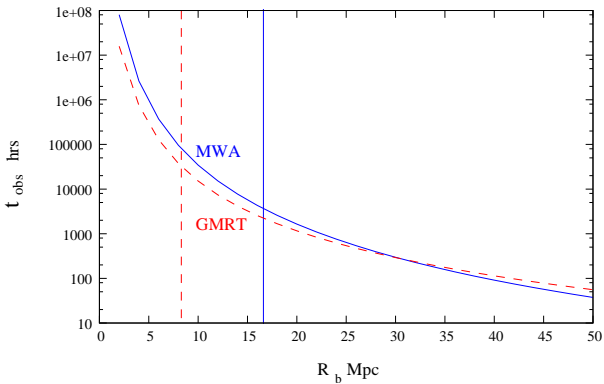}
\caption{The observing time $t_{obs}$  that would be required for a $3
  \, \sigma$ detection of  a bubble of radius $R_b$ provided it is at
  the center of the 
field of view. The vertical lines shows the lower limit (due to HI
fluctuations) where a detection will be possible ($R_b=8 \, {\rm Mpc}$
for GMRT and $R_b=16 \, {\rm Mpc}$ for MWA).}
\label{fig:R_Tobs}
\end{figure}

\begin{figure*}
\includegraphics[width=0.9\textwidth]{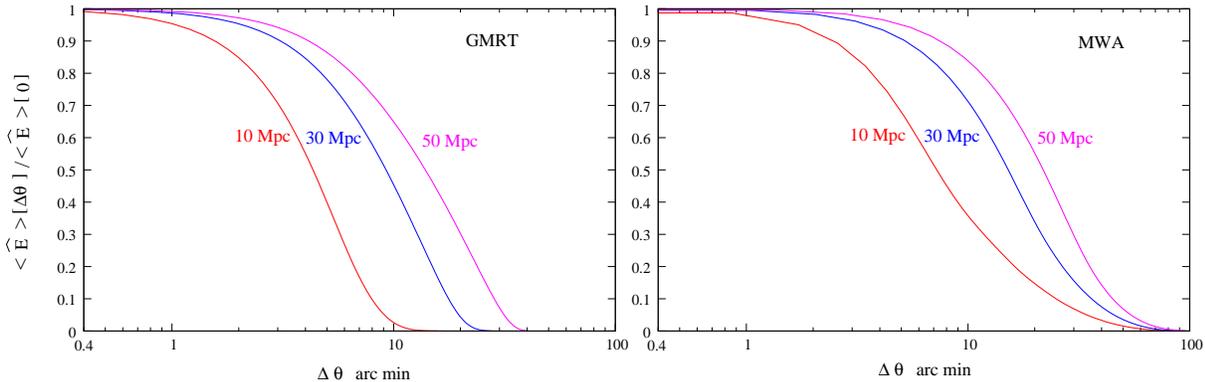}
\caption{The Overlap between the signal and the filter when there is a
  mismatch $\Delta \theta$ between the centers of the bubble and the
  filter for GMRT (left) and MWA (right). The results are shown for 
different bubble sizes.}
\label{fig:dtheta}
\end{figure*}

At  both the GMRT and the MWA, for $100 \, {\rm hrs}$ of observation,
the noise is larger than the signal for bubble  size $R_b \le 40 
\,{\rm   Mpc}$.  At the other extreme, for an  integration time of
$10^5 \, {\rm hrs}$   the noise is  below the signal for $R_b > 6
\,{\rm  Mpc}$ for the GMRT and $R_b > 8 \,{\rm   Mpc}$ for the MWA. 
The  foreground contribution  turns out to be smaller  than the
signal for the entire range of bubble sizes that we have considered, 
thus justifying our choice of filters. 
Note that 
Filter II is more efficient in foreground subtraction, but it requires
prior knowledge about the frequency dependence. 
For both the  filters the foreground  removal is more effective at the
GMRT 
than the MWA  because of the frequency dependence of
$\rho_N(\u,\nu)$. The  assumption  that this  is proportional to
$\lambda^2$ is valid only when $\rho(\u,\nu)$ is independent of $U$,
which, as we have discussed, is true  for a large $U$ range at the
GMRT. The $\lambda$ dependence is much more complicated at the MWA,
but we have not considered such details here as the foreground
contribution is 
anyway smaller than the signal. It should also be noted that the foreground
contribution increases at small baselines (eq. \ref{eq:fg}),  
and  is very sensitive to
the  smallest value of $U$ which we set at $U=20$ for our
calculations. Here it must be noted that our results  are valid only under the
assumption that the
foregrounds have a smooth frequency dependence. A slight deviation from this
and 
the signal will be swamped by the foregrounds. Also note that this filtering
method is effective only for the detection of the bubbles and not for
the statistical HI fluctuations signal. 

The contribution from the HI fluctuations impose a lower limit on the
size of the bubble which can be detected. However long be the
observing time,  it will not be
possible to detect bubbles of size  $R_b<8 \, {\rm Mpc}$ using  the
GMRT and  size $R_b<16 \, {\rm Mpc}$ using the MWA. The HI fluctuation
contribution increases at small baselines. The problem is particularly
severe at MWA because of  the  dense sampling of the small baselines
and the  very large field of view. We note that the MWA is being
designed with the detection  of the statistical HI
fluctuation  signal in mind, and hence it is not surprising  that this
contribution is quite large. 
For both  telescopes it 
 may be possible to reduce this  component by cutting off the filter
 at small baselines. We have not explored this possibility in this work
because  the enormous
 observing times required to detect such small bubbles makes
 it unfeasible with the GMRT or MWA.

Figure \ref{fig:R_Tobs} shows the observation time that would be
required to detect bubbles of different sizes using Filter I for 
GMRT and the MWA. Note that the observing time shown here refers to a
$3 \, \sigma$ detection which is possibly adequate for targeted
searches centered on observed quasar position. A more stringent
detection criteria at the $5 \, \sigma$ level would be apropriate for
a  blind search. The observing time would go up by a factor of $3$ for
a $5 \, \sigma$ detection.  
The
observing time is similar for Filter II and hence we do not show this 
separately. 
In calculating the observing time we have only taken into
account the noise contribution as the other contributions do not change
with time. The  value of $R_b$ below which a detection  is not
possible due to the HI fluctuations is shown by vertical lines 
for both telescopes. 
We see that with $100 \,{\rm hrs}$ of observation both the telescopes
will be able to detect bubbles with $R_b>40\, {\rm   Mpc}$ while 
bubbles with  $R_b>22 {\rm  Mpc}$ can be detected with $1000 \,{\rm
  hrs}$ of observation.

The possibility of detecting a bubble is less when the bubble
centre does not coincide with the centre of the field of view.
In fact, the SNR  falls as $e^{-\theta_c^2/\theta_0^2}$ if the
bubble center is shifted  away by $\theta_c$ from the center of the
field of view and the filter is  also shifted so that its center
coincides with 
that of the bubble.  There will be a corresponding increase 
$t_{obs} \propto e^{2 \theta_c^2/\theta_0^2}$ in the observing time
required to detect the bubble. It will be possible to detect bubbles
only if they are located near the center of the field of view
($\theta_c \ll \theta_0$), and the required observing time increases
rapidly with $\theta_c$ for off-centered bubbles.

When searching for bubbles in a particular observation it will be
necessary to consider filters corresponding to all possible  value of 
$R_b$, $\nu_c$ and $\th_c$. A possible strategy would be to
search at  a discrete set of values in the range of $R_b$, $\nu_c$ and
$\th_c$ values where a detection is feasible. 
The crucial issue here would be the choice of the sampling density
 so that we do not miss out an ionized bubble whose
parameters  do not exactly  coincide with any of the values in  
the discrete set  and  lie somewhere in between.  To illustrate this
we discuss the considerations for choosing and optimal value of 
$\Delta \theta_c$ the sampling interval for $\th_c$. We use 
$\langle \hat{E} \rangle [\Delta  \theta]$   to denote the expectation
value of the estimator when there is a  mismatch $\Delta \theta$
between the  centers   of the bubble and the filter. The ratio
${\rm   Overlap}=\langle \hat{E} \rangle[\Delta \theta]/\langle 
\hat{E}\rangle[0]$, shown in   Figure \ref{fig:dtheta} for 
GMRT (left panel) and MWA (right panel), 
 quantifies the overlap between the signal and the
filter   as $\Delta \theta$ is varied. We see that the choice of
$\Delta \theta$ would depend on the size of the bubble we are trying to
detect and  it would be smaller for the GMRT as compared to the
MWA. Permitting the Overlap to drop to $0.9$ at the middle of the sampling
interval, we find that it is $8^{'}$ at 
the GMRT and $20^{'}$ at  the MWA for  $R_b=50 \, {\rm Mpc}$.  

\begin{figure}
\includegraphics[width=0.45\textwidth]{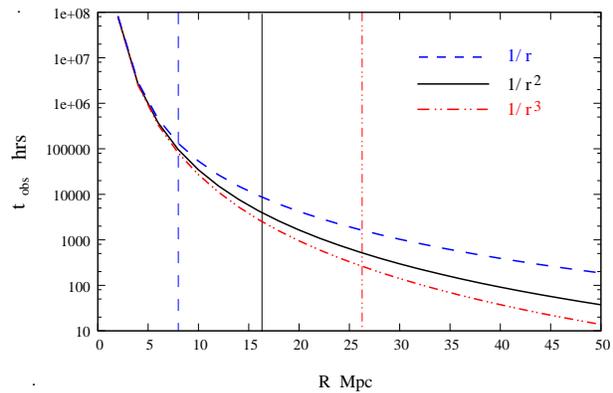}
\caption{Same as the Figure \ref{fig:R_Tobs} considering three
  different antenna distributions 
$\rho_{ant}(r)\propto 1/r$,\, $1/r^2$,\, $1/r^3$ \, for the MWA.}
\label{fig:R_Tobs_rho}
\end{figure}

The MWA is  yet to be  constructed, and it may be possible
that  an  antennae distribution different from
$\rho_{ant}(r)\propto 1/r^2,\,$ may improve the prospects of detecting
HII bubbles.  We have tried out $\rho_{ant}(r)\propto  1/r \, {\rm
  and} \, 1/r^3$ for which the results are shown in 
Figure~\ref{fig:R_Tobs_rho}.  We find that the required  
integration time falls considerably for the $1/r^3$ distribution
whereas the opposite occurs for $1/r$. 
For example, for $R_b=50 \,{\rm Mpc}$ the integration time increases
by  $5$ times for $1/r$ and decreases by $3$ times for  $1/r^3$ as
compared to $1/r^2$. Based on this we expect the integration time to
come down if the antenna distribution is made steeper, but this occurs
at the expense of increasing the HI fluctuations and the
foregrounds. We note that for the $1/r^3$ distribution the foreground
contribution is more than the signal, but it may be possible to
overcome this by modifying the filter. The increase in the HI
fluctuations is inevitable, and it restricts the smallest bubble that
can be detected to $R=26 \, {\rm MPc}$ for $1/r^3$. In summary, the
$1/r^2$ distribution appears to be a good compromise between reducing
the integration time and increasing the HI fluctuations and
foregrounds.

Finally we examine some of the assumptions made in this work.
First, the Fourier
relation between the specific intensity and the visibilities
(eq. \ref{eq:1}) will be valid only near the center of the field of
view and  full three dimensional wide-field imaging is needed away
from the center. As the feasibility  of detecting a bubble  away from the 
center falls rapidly, we do not expect the wide-field effects to be very
important. Further, these effects are most significant at large
baselines whereas  most of the signal from ionized bubbles is in the
small  baselines.

Inhomogeneities in the IGM will affect the propagation of ionization
fronts, and the ionized bubbles are not expected to be exactly
spherical \citep{wyithe05}. This will cause a mismatch between the
signal and the filter which in turn will degrade the SNR. In addition
to this, in future we plan to address a variety of other issues like
considering different observing frequencies and  making predictions
for the other upcoming  telescopes. 

Terrestrial signals  from television, FM radio,
satellites, mobile communication etc., collectively referred to as
RFI,  fall in the same frequency band as the redshifted $21 \rm cm$ signal
from the reionization epoch.  These are
expected to be much stronger than the expected $21 {\rm cm}$ signal, and
it is necessary to quantify and characterize the RFI. 
 Recently \citet{bowman07} have  characterized   the RFI for the 
 MWA  site on the frequency range $80$ to $300 \rm MHz$. They find an
 excellent  RFI environment except for a few channels which are
 dominated by satellite  communication signal.  The impact  of RFI on
 detecting ionized bubbles is an important issue which we plan to
 address in future. 

The effect of polarization leakage is another issue we postpone for
future work. This could cause polarization structures on the sky to
appear as frequency dependent ripples in the foregrounds intensity
. This could be particularly severe for the MWA.

\section*{Acknowledgment}
KKD would like to thank Sk. Saiyad Ali, Prasun Dutta, Ravi
Subramanyam, Uday Shankar and TRC would like 
to thank Ayesha Begum for useful discussions. We thank the anonymous
referee for useful comments.
KKD is supported by a senior research fellowship of Council of 
Scientific and Industrial Research (CSIR), India.

\appendix

\section{Relation between visibility-visibility correlation and MAPS}
\label{sec:vis}
In this appendix we give the calculations for expressing the 
two visibility correlation in terms of the Multi-frequency 
angular power spectrum (MAPS). We can write the visibility 
$V(\u,\nu)$ as a two-dimensional Fourier transform 
of the brightness temperature $T(\th, \nu)$ [see equation (\ref{eq:1})]
\be
V(\u,\nu)=\left(\frac{\del B}{\del T}\right)_{\nu}
\int d^2 \theta A(\th, \nu) T(\th, \nu)
e^{ 2\pi \imath \th \cdot \u}
\e
where $(\del B/\del T)_{\nu}$ is the conversion factor from 
temperature to specific intensity and $A(\th, \nu)$ is the 
beam pattern of the individual antenna. The visibility-visibility correlation
is then given by
\begin{eqnarray}
\langle V(\u_1,\nu_1) V(\u_2 ,\nu_2) \rangle \!\!\!\!\!&=& \!\!\!\!\!
\left(\frac{\del B}{\del T}\right)_{\nu_1}
\left(\frac{\del B}{\del T}\right)_{\nu_2}
\nonumber\\
\!\!\!\!\!&\times&\!\!\!\!\!
\int d^2 \theta \int d^2 \theta'
A(\th,\nu_1) A(\th',\nu_2)
\nonumber\\
\!\!\!\!\!&\times&\!\!\!\!\!
\langle T(\th, \nu_1) T(\th', \nu_2) \rangle
e^{2\pi \imath (\th \cdot \u_1 + \th' \cdot \u_2)}
\nonumber\\
\label{eq:vis-vis}
\end{eqnarray}
The correlation function for the temperature fluctuations on the sky
would simply be the two-dimensional Fourier transform of the MAPS 
$C_{2 \pi U}(\nu_1,\nu_2)$
\be
\langle T(\th, \nu_1) T(\th', \nu_2) \rangle = 
\int d^2 U ~ C_{2 \pi U}(\nu_1,\nu_2) 
e^{-2\pi \imath (\th - \th') \cdot \u}
\e
Using the above equation in equation in (\ref{eq:vis-vis}), we obtain
\begin{eqnarray}
\langle V(\u_1,\nu_1) V(\u_2 ,\nu_2) \rangle \!\!\!\!\!&=& \!\!\!\!\!
\left(\frac{\del B}{\del T}\right)_{\nu_1}
\left(\frac{\del B}{\del T}\right)_{\nu_2}
\nline
\!\!\!\!\!&\times&\!\!\!\!\!
\int d^2 U ~ C_{2 \pi U}(\nu_1,\nu_2) 
\nonumber\\
\!\!\!\!\!&\times&\!\!\!\!\!
\tilde{A}(\u_1-\u, \nu_1) \tilde{A}(\u_2+\u, \nu_2) 
\end{eqnarray}
where $\tilde{A}(\u, \nu)$ is the Fourier transform of the 
beam pattern $A(\th, \nu)$. If the beam pattern is assumed to be Gaussian
$A(\th, \nu) = e^{-\theta^2/\theta_0^2}$, the 
Fourier transform too is given by a Gaussian function
\be
\tilde{A}(\u, \nu) = \pi \theta_0^2 e^{-\pi^2 U^2 \theta_0^2}
\e
Hence, the visibility correlation becomes
\begin{eqnarray}
\langle V(\u_1,\nu_1) V(\u_2 ,\nu_2) \rangle \!\!\!\!\!&=& \!\!\!\!\!
\left(\frac{\del B}{\del T}\right)_{\nu_1}
\left(\frac{\del B}{\del T}\right)_{\nu_2}
\pi^2 \theta_1^2 \theta_2^2
\nline
\!\!\!\!\!&\times&\!\!\!\!\!
\int d^2 U ~ C_{2 \pi U}(\nu_1,\nu_2) 
\nonumber\\
\!\!\!\!\!&\times&\!\!\!\!\!
e^{-\pi^2[(\u_1 - \u)^2 \theta_1^2 + (\u_2 + \u)^2 \theta_2^2]}
\end{eqnarray}
where $\theta_1$ and
$\theta_2$ are the values of $\theta_0$ at $\nu_1$ and $\nu_2$
respectively. Now, since the two Gaussian functions in the above
equation is peaked around different values of $\u$, the 
integrand will have a non-zero contribution only
when $|\u_1 + \u_2| < (\pi ~ {\rm max}[\theta_1, \theta_2])^{-1}$.
In case the typical baselines are much larger than the quantity
$(\pi ~ {\rm max}[\theta_1, \theta_2])^{-1}$, 
the integral above can be well approximated as  being  non-zero only
when $\u_1 = -\u_2$. Then 
\begin{eqnarray}
\langle V(\u_1,\nu_1) V(\u_2 ,\nu_2) \rangle \!\!\!\!\!&\approx& \!\!\!\!\!
\delta_{\u_1,-\u_2} 
\left(\frac{\del B}{\del T}\right)_{\nu_1}
\left(\frac{\del B}{\del T}\right)_{\nu_2}
\nline
\!\!\!\!\!&\times&\!\!\!\!\!
\pi^2 \theta_1^2 \theta_2^2 C_{2 \pi U_1}(\nu_1,\nu_2)
\nline
\!\!\!\!\!&\times&\!\!\!\!\! 
\int d^2 U
e^{-\pi^2[(\u_1 - \u)^2 (\theta_1^2 + \theta_2^2)]}
\nonumber\\
\!\!\!\!\!&=& \!\!\!\!\!
\delta_{\u_1,-\u_2} \pi \left(\frac{\theta_1^2 \theta_2^2}{\theta_1^2
  + \theta_2^2}\right)
 \left(\frac{\del B}{\del T}\right)_{\nu_1}\left(\frac{\del B}{\del
   T}\right)_{\nu_2}
\nline
\!\!\!\!\!&\times&\!\!\!\!\! 
C_{2 \pi U_1}(\nu_1,\nu_2) 
\end{eqnarray}
which is what has been used in equation (\ref{eq:11}).

In the continuum limit, the Gaussian $\tilde{A}(\u, \nu)$ can be approximated
by a delta function, i.e., $\tilde{A}(\u, \nu) \approx \delta^{(2)}_D(\u)$
(which corresponds to the limit $\theta_0 \to \infty$);
the visibility-visibility correlation is then given as 
\begin{eqnarray}
\langle V(\u_1,\nu_1) V(\u_2 ,\nu_2) \rangle
= \delta^{(2)}_D(\u_1 + \u_2) \left(\frac{\del B}{\del
  T}\right)_{\nu_1} \left(\frac{\del B}{\del T}\right)_{\nu_2}
\nline 
\times C_{2 \pi U_1}(\nu_1,\nu_2) 
\end{eqnarray}
which corresponds to equation (\ref{eq:21}) in the main text.

\end{document}